\documentclass[12pt,preprint]{aastex}
 \pdfoutput=1

\usepackage{color}

\slugcomment{Not to appear in Nonlearned J., 45.}

\shorttitle{The Nature of Coronal Dimming} \shortauthors{Cheng et
al.}

\begin{document}


\title{THE NATURE OF CME-FLARE ASSOCIATED CORONAL DIMMING}


\author{J. X. Cheng\altaffilmark{1,2}, J. Qiu \altaffilmark{3}}
\affil{$^1$Key Laboratory of Planetary Sciences, Shanghai Astronomical Observatory, Shanghai 200030, China} \email{chengjx@shao.ac.cn}

\affil{$^2$Key Laboratory of Solar Activity, National Astronomical
Observatories, Chinese Academy of Sciences, Beijing 100012}
\affil{$^3$Department of Physics, Montana State University,
Bozeman MT 59717-3840, USA
}

\begin{abstract}

Coronal mass ejections (CMEs) are often accompanied by coronal
dimming evident in extreme ultraviolet (EUV) and soft X-ray
observations. The locations of dimming are sometimes considered to
map footpoints of the erupting flux rope. As the emitting material
expands in the corona, the decreased plasma density leads to
reduced emission observed in spectral and irradiance measurements.
Therefore, signatures of dimming may reflect properties
of CMEs in the early phase of its eruption. In this study, we
analyze the event of flare, CME, and coronal
dimming on December 26, 2011. We use the data
from the Atmospheric Imaging Assembly (AIA) on
Solar Dynamics Observatories (SDO) for disk observations
of the dimming, and analyze images taken by EUVI, COR1, and COR2
onboard the Solar Terrestrial Relations Observatories to obtain
the height and velocity of the associated CMEs observed at the limb.
We also measure magnetic reconnection rate from flare observations.
Dimming occurs in a few locations next to the flare ribbons, and it
is observed in multiple EUV passbands. Rapid dimming starts after the
onset of fast reconnection and CME acceleration, and its evolution
well tracks the CME height and flare reconnection. The spatial
distribution of dimming exhibits cores of
deep dimming with a rapid growth, and their light curves are approximately
linearly scaled with the CME height profile. From the dimming analysis,
we infer the process of the CME expansion, and estimate properties of the CME.

\end{abstract}

\keywords{Sun: coronal mass ejections --- Sun: flares---Sun: EUV
dimming---Sun: magnetic reconnection}

\section{INTRODUCTION}
 Flares and coronal mass ejections are manifestations of the most energetic large-scale solar eruptions and often occur together. In addition, coronal mass ejections are the main source of strong geomagnetic storms. Geomagnetic storms can be also caused by corotating interaction regions (CIRs) \citep{Balogh1999} when hight speed solar streams emanating from coronal holes interact with streams of lower speed. During the eruptive events, transient coronal holes, or coronal dimmings,
 are often observed \citep{thompson00,harrison03,zhukov04}.
Coronal dimming was first observed in Skylab data and characterized as transient
coronal holes \citep{rust76,rust83}. Subsequently, similar
observations have been analyzed to study the relationship of dimming
with CMEs, flares and other associated phenomena
\citep{hudson96,sterling97,harrison03,zhukov04}. By these series
of studies, dimmings are usually interpreted as mass depletion
due to the loss or rapid expansion of the overlying corona
\citep{hudson98,harrison00,zhukov04}. This interpretation is
supported by imaging observations of simultaneous and co-spatial dimming
in several coronal lines \citep[e.g.,][]{zarro99,sterling00}, as well as
spectroscopic observations \citep{harra01,tian12}. Although CMEs
are also observed to occur without dimming, \citet{reinard09}
found that non-dimming CMEs all have speeds of less than 800 km
s$^{-1}$, suggesting a more intimate connection between fast CMEs
and dimming properties. \citet{krista13} found further
correlations between the magnitudes of dimmings and flares, and the CME mass by
studying variations between recurring eruptions and dimmings.

Coronal dimming can be produced by various processes, although the
main contributor is mass depletion. As summarized by
\citet{mason14}, several different mechanisms have been proposed to explain
coronal dimming. (1) Mass-loss dimming. The mass-loss dimming
is produced by ejection of emitting plasma \citep{harrison00,harra01},
which causes darkening of the areas in and near the erupting active region.
\citet{harrison03} showed that coronal mass estimated from dimming signatures
can account for a large percentage of the CME mass. Additional
observations from the Hinode and SDO spacecraft have confirmed that
the coronal plasma density decreases with accompanying outflows
\citep{attril10,harra11,tian12}. (2) Thermal
Dimming. This is due to heating of coronal plasmas to higher temperatures,
so that heated areas appear dark in EUV or SXR channels sensitive
to lower temperatures, but these areas are brightened simultaneously in channels sensitive to higher temperatures.
(3) Obscuration Dimming. Here the physical process that results in apparent dimming
is an absorptive material (e.g., a filament) moving between bright materials
(e.g., flare arcade) and the observer. (4) Wave Dimming. One of the simplest explanations
of the observations is that plasma is compressed as a longitudinal
wave passes through the medium.  (5) Doppler Dimming. If the plasma
is moving with a line-of-sight motion component, the coronal emission lines
are Doppler-shifted, which may cause reduction of the total counts observed
in a filtergram with a fixed and narrow band-pass, leading to an apparent
dimming in the data.

Among these mechanisms, the mass-loss dimming is considered to be
the main process of coronal dimming, and
this scenario is supported by many recent studies
\citep{sterling97,reinard08,reinard09,aschwanden09}.
Post-flare dimmings detected by the EVE instrument on SDO are used
to estimate the CME mass \citep{hudson14}. Based upon earlier
X-ray observations, \citet{hudson14} interpret these dimmings as the result of
CME mass ejection from the low corona. They estimate the mass
from the dimming by deriving a best pre-event
temperature and emission measure in the dimmed region from EVE,
and a source volume from AIA images. \citet{tian12} suggest that
dimming is mainly an effect of density decrease rather than
temperature change by plasma diagnostics of the dimming region.
The mass losses in dimming regions have been estimated from two
different methods, and they make 20\% - 60\% of the masses of the
associated CMEs, suggesting that a significant part of the CME
mass indeed comes from the region where dimming occurs
subsequently. The mass flux carried by the outflows has also been
estimated from observations.

Considering that dimmings map the foot-points of erupting flux ropes,
magnetic flux in dimming regions is measured to estimate the total flux
in the flux rope \citep{Webb2005, Qiu2007}.
Detailed analysis of dimming observations also demonstrate evolution
of dimming in terms of both the area of dimming regions and the depth of dimming.
\citet{Miklenic2011} reported a twin-dimming event near the feet of an erupting
sigmoid CME, both of which also expand in area. They interpret such evolution
as due to interaction of erupting flux rope with overlying field lines.
\citet{Downs2015} conducted comprehensive, data-driven numerical simulations
and topological analysis of coronal dimmings associated with an erupting flux rope,
and demonstrated three types of coronal dimming and their evolution. The first two types
of dimming map the feet of the erupting rope, or of overlying field lines reconnecting with the
rope and subsequently entrained into the erupting rope. The third
type of dimming maps the feet of the arcade adjacent to the rope, which
does not erupt. These studies suggest that coronal dimming evolution is related
to the reconstruction of magnetic structures during the flux rope eruption. Therefore,
analysis of dimming observations can provide some measurements of CME
properties.

It is widely accepted that energy stored in the magnetic field is
released to the solar atmosphere by magnetic reconnection.
\citet{Subramanian2007} showed that on average, the
magnetic energy contained in CMEs provides at least 74 \% of what
is required for their propagation from the Sun to the Earth. A
large majority of CMEs observed with coronagraphs are now
confirmed to possess a flux rope morphology \citep[e.g.,][]{Vourlidas2013,Zhang2013}. Many studies are focused on figuring
out the CME evolution from a 3D viewpoint
\citep{subramanian14,thernisien09,gopalswamy12}. They study the
CME evolution based on different models and for varied purposes.
Although there has been some observational evidence for the fact
that some flux rope CMEs expand in a self-similar manner in the
coronagraph field of view
\citep{poomvises10,kilpua12,colaninno13}, there are still some
events which can not be described in a self-similar expansion model.
Self-similarity has been invoked in a number of theories of
CME propagation \citep[e.g.,][]{Kumar1996,Demoulin2009,Wang2009,Olmedo2010}. Other models, such
as full ice-cream cone, partial ice-cream cone, and flat cone models,
are used to describe propagations of some CMEs
\citep{gopalswamy12}.

In this paper, we analyze observations of a coronal dimming event associated with
a halo CME and a two-ribbon flare on December 26, 2011. Observations are obtained
by SDO and STEREO. We find two dimming regions at opposite ends of two flare ribbons, which probably
map the feet of the erupting CME, and also a dimming band along a flare ribbon, which
may map the feet of the magnetic arcade overlying the erupting CME. Our analysis shows
that the time profiles of coronal dimming at various locations well track the CME height
evolution and magnetic reconnection. The spatial distribution of dimming exhibits cores
of deep dimming, whose evolution appears to agree with the prediction by a one-dimensional
isothermal CME expansion model in the early phase of the eruption. The paper is organized
as follows. Observations are described in Section 2, followed by data analysis and results
in Section 3. Discussions and conclusions are given in Section 4.

\section{OBSERVATIONS }

\begin{figure}
\epsscale{0.8} \plotone{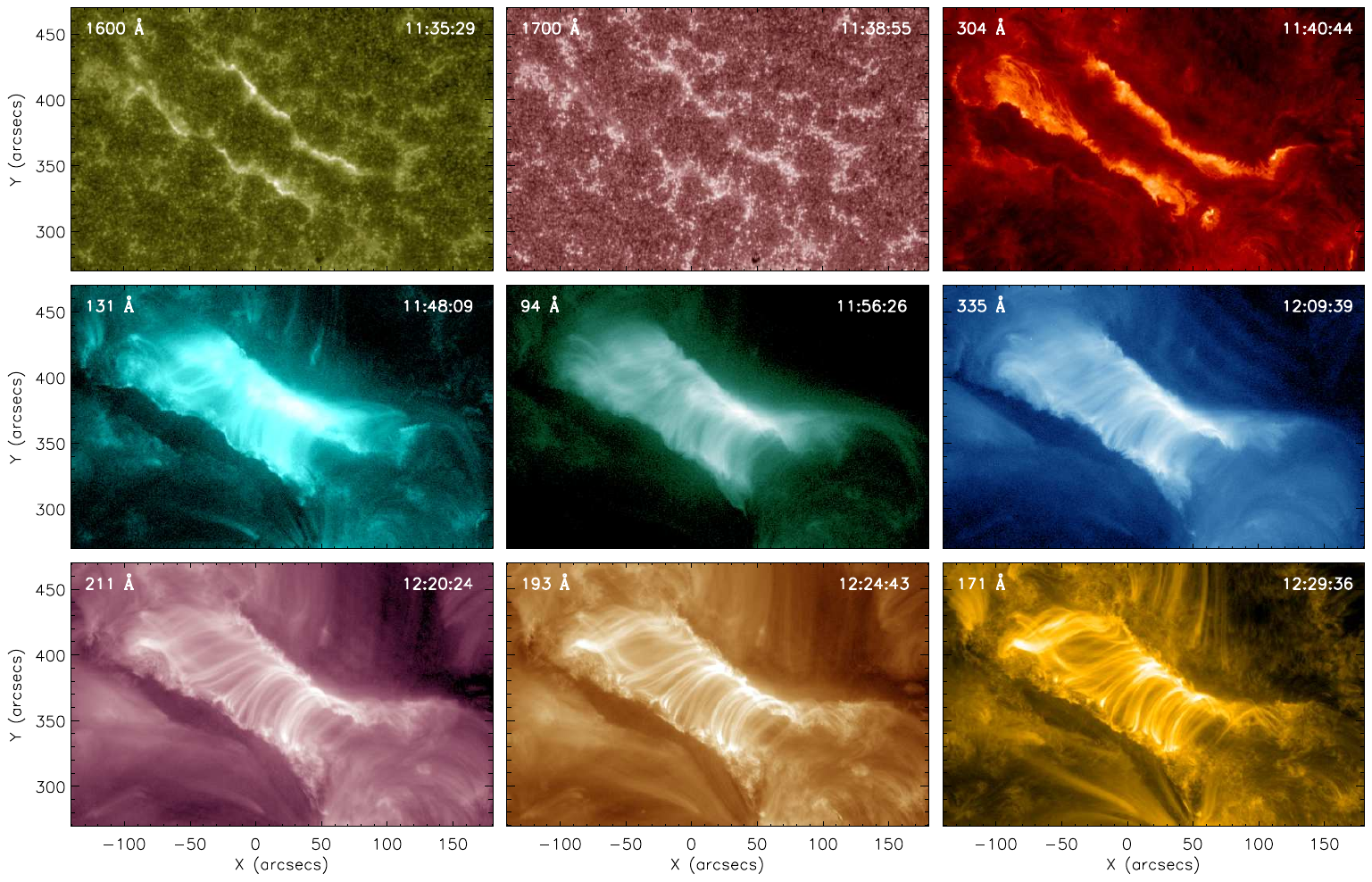}
 \caption{AIA images taken at peak times of given passbands
for the 2011 December 26 event. The two-ribbon flare is
located at the solar disk center and ribbon brightening spreads along
the polarity inversion line from southwest to northeast. Post flare loops are clearly seen
in multiple EUV channels. \label{fig1}}
\end{figure}

In this study, we analyze an eruptive event on 2011 December 26. A C5.7 flare takes
place in AR 11384 at N17 W02. It is associated with a fast halo CME as well as coronal dimming.
In the event, the flare and coronal dimming take place near the disk center, and are observed by
the Solar Dynamics Observatory's \citep[SDO;][]{pesnell12} Atmospheric Imaging Assembly
\citep[AIA;][; see Figure~\ref{fig1}]{lemen12}. AIA is a full-disk imager observing in seven EUV and two UV bandpasses
with a moderate cadence (12s in EUV channels and 24s in UV channels) and high resolution
(0.6 $''$ pixel$^{-1}$ spatial scale). The Helioseismic and Magnetic Imager \citep[HMI;][]{schou12}
onboard SDO measures magnetic fields in the Sun's photosphere. HMI observes the full solar
disk at 6173 \AA\ with a spatial scale of 0.5\arcsec\ per pixel. Magnetic reconnection
flux is measured during the flare by summing up magnetic flux encompassed by flare ribbons \citep{qiu2002,qiu2004}.

\begin{figure}
\epsscale{0.8} \plotone{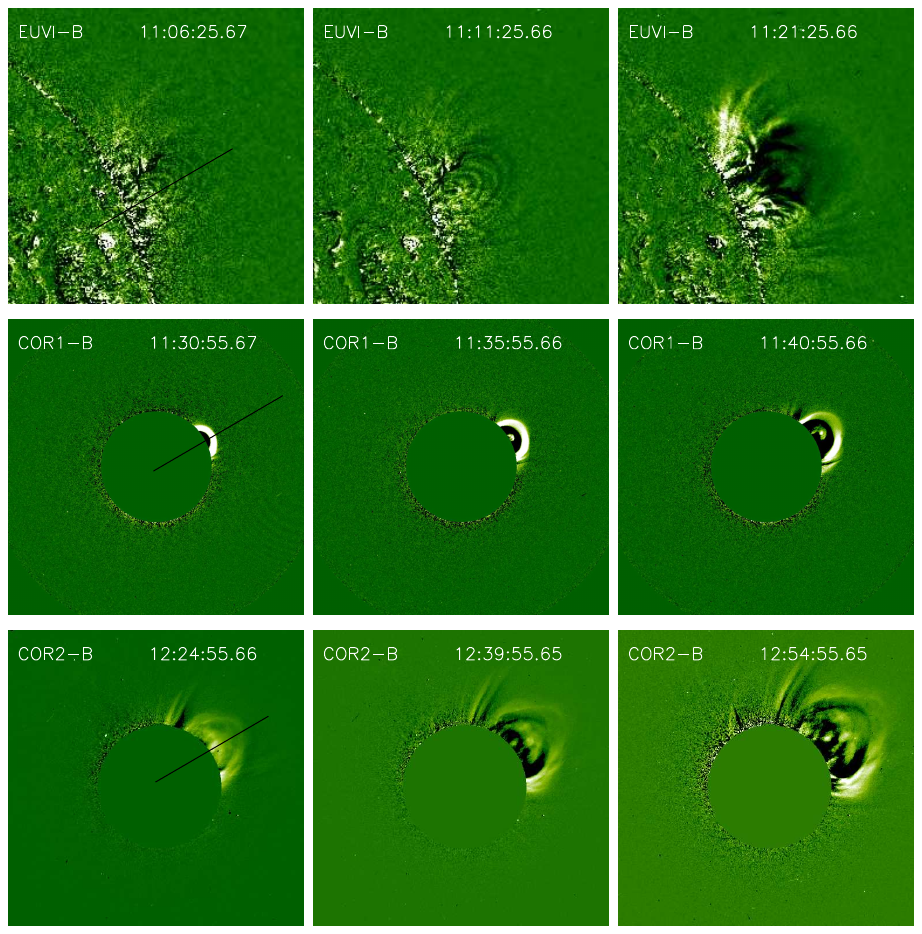} \caption{CME
evolution observed by STEREO EUVI, COR1 and COR2.  A very bright
core is clearly seen from COR1.\label{fig2}}
\end{figure}

The CME that is associated with the flare is observed by the
Solar Terrestrial Relations Observatory (STEREO). STEREO is a pair of
counter-rotating Sun satellites at roughly 1 astronomical unit
(AU) from the Sun.  Extreme UltraViolet Imager \citep[EUVI;][]{wuelser04} observes
the full solar disk in 304\AA, 171\AA, 195\AA, 284\AA, in field of view to
1.7 $R_\odot$ with a spatial scale of 1.6\arcsec pixel$^{-1}$. The inner (COR1)
and outer (COR2) \citep[SECCHI;][]{howard08} white light coronagraphs provide
images from 1.5 to 4$R_\odot$ and out to 15$R_\odot$, respectively. In 2011,
the STEREO A and B are located at positions about $\pm 90$ degrees from the SDO,
providing a very good opportunity to simultaneously observe flares, coronal dimmings,
and magnetic fields on the Sun's disk and CME propagation from the limb. In this study, we use observations
by EUVI, COR1, and COR2 on STEREO to track the height of the CME (Figure~\ref{fig2}),
and examine its evolution together with the flare and coronal dimming.

\section{ DATA ANALYSIS \& RESULTS}
\subsection{Flare, CME, and Magnetic Reconnection}
\begin{figure}
\epsscale{0.8} \plotone{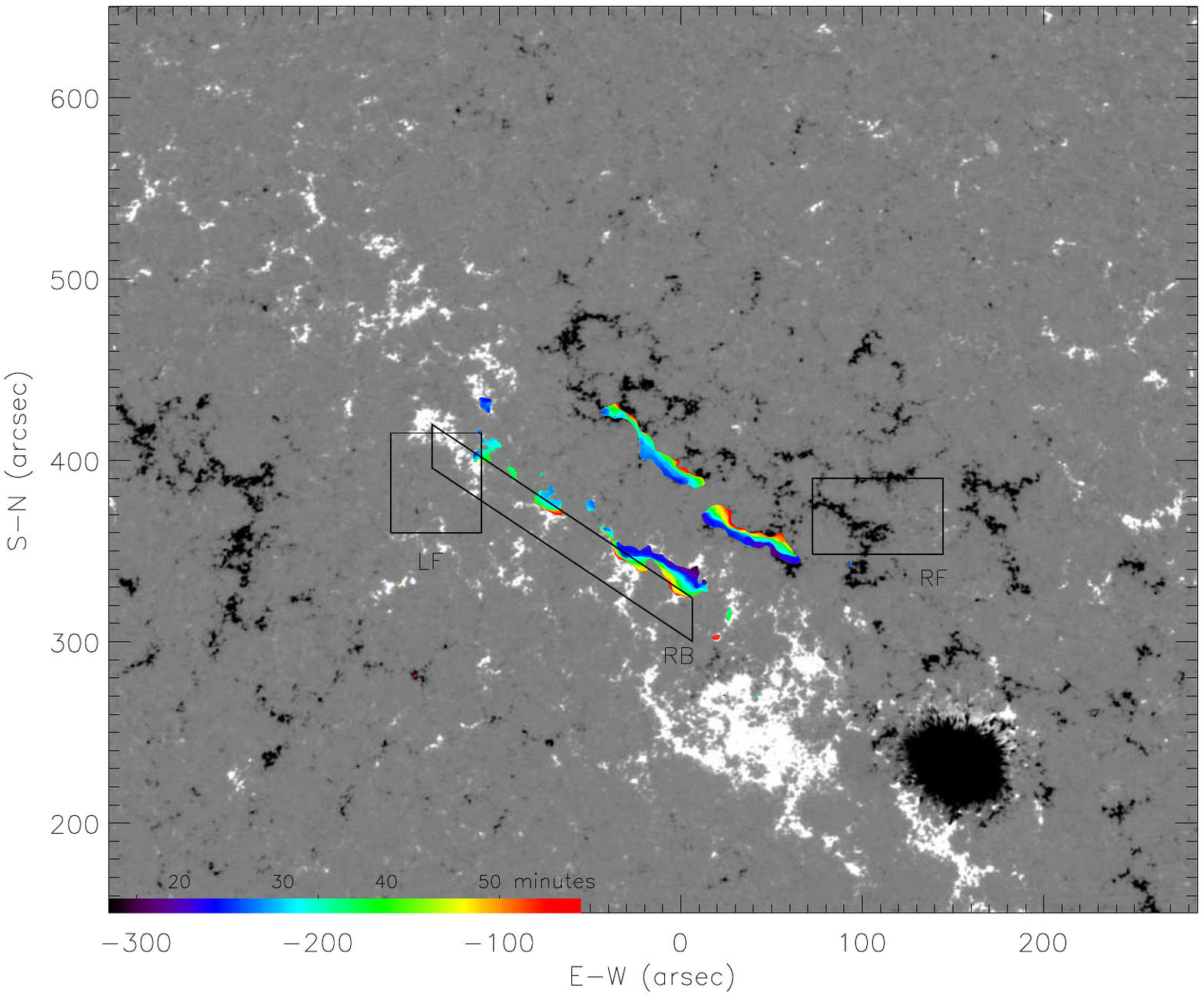} \caption{Longitudinal magnetogram of the flaring region from SDO/HMI.
Colors indicate the time sequence of flare ribbon brightening after 11:00~UT, and the three boxes
indicate regions of interest where dimming is primarily located. Dimming pixels are selected
from these regions for further analysis in the following figures.
 \label{fig3}}
\end{figure}

\begin{figure}
\epsscale{0.8} \plotone{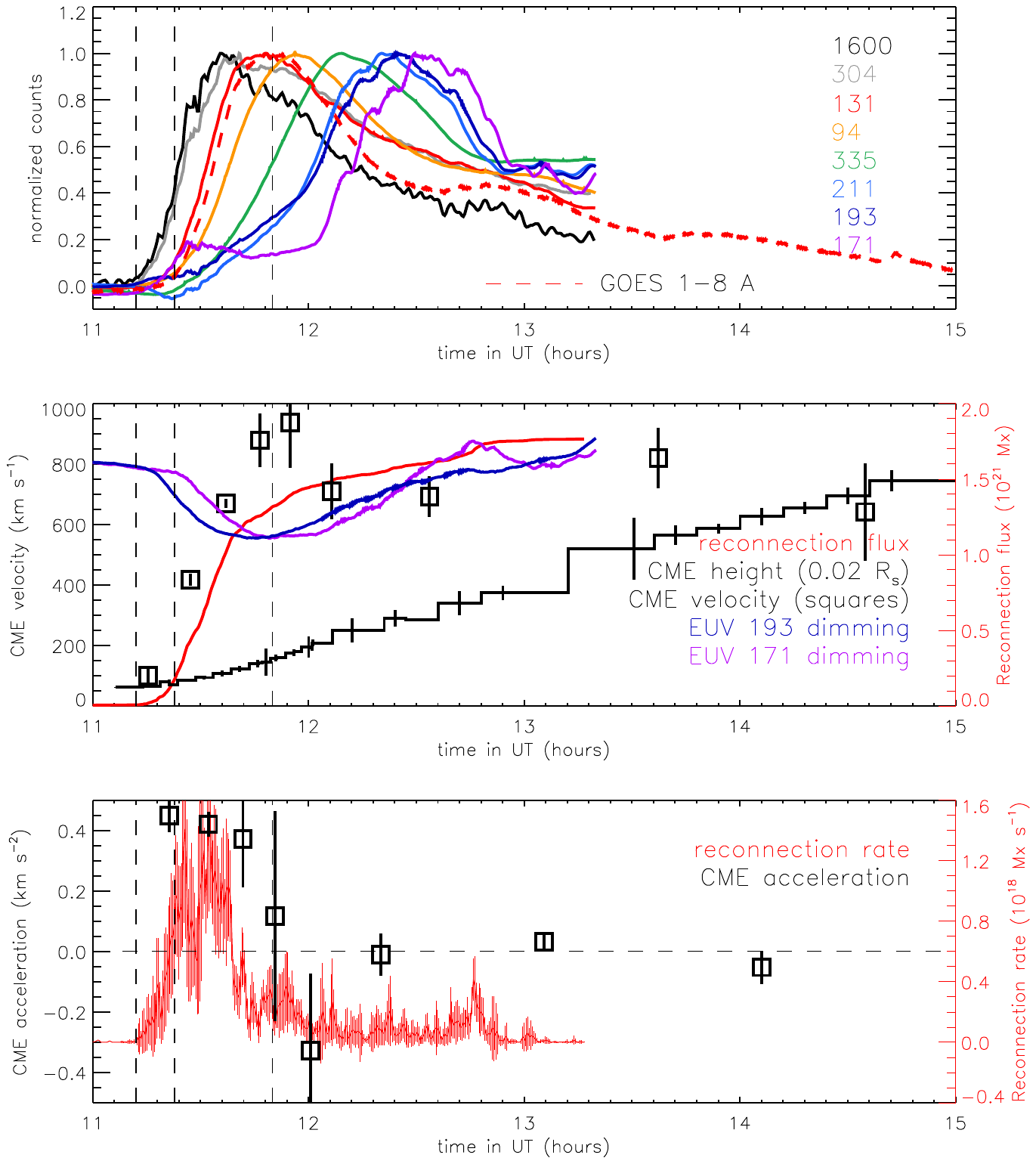}
\caption{Upper: Integrated EUV counts over the flaring region shown in Figure 1, and GOES SXR light curve.
All light curves are normalized to the maximum.
Three vertical lines indicated the times of start of flare ribbon brightening, and rise and peak of GOES SXR emission.
Middle: CME height, derived velocity, magnetic reconnection flux,
and total EUV counts in 193 and 171 bands of dimming pixels in the active region.
Bottom: Derived CME acceleration and magnetic reconnection rate. \label{fig4}}
\end{figure}

The eruptive event takes place near the disk center. It produces a C-5.7 flare.
According to the GOES catalog \footnote{\url{ftp://ftp.ngdc.noaa.gov/STP/space-weather/solar-data/solar-features/solar-flares/x-rays/goes/2011/}} , the soft X-ray (1-8 \AA) emission of the flare
starts to rise at 11:23~UT, peaks at 11:50~UT, and lasts till 12:18~UT. The flare is
well observed by SDO/AIA. Figure~\ref{fig1} gives an overview of the flare
in nine UV/EUV wavelengths. Each image is taken at the peak time of the total
emission in the given bandpass.
The flare is a very typical two-ribbon flare with brightening of
the two ribbons visible in multiple passbands. The ribbon
brightening quickly spreads from southwest to northeast along the magnetic
polarity inversion line (PIL), and then expands outward away from the PIL.
Such a time sequence of ribbon brightening is displayed in Figure~\ref{fig3}, superimposed
on a longitudinal magnetogram obtained by SDO/HMI before the flare onset.
Subsequently, post flare loops are clearly seen in the EUV channels. The light curves observed by AIA
for the flaring region are displayed in the top panel in Figure~\ref{fig4}.
The figure shows that UV and EUV emissions at the flare ribbons, particularly in the
1600 and 304\AA~ bands,
start to rise at around 11:10~UT, 10 minutes before the rise of the GOES SXR emission.
In the plots, the vertical dashed lines indicate the start of the flare ribbon brightening, as well as the start
and peak times of the soft X-ray emission observed in GOES, respectively.

According to the standard flare model \citep{carmichael64,sturrock66,hirayama74,kopp76},
expansion of flare ribbons indicates progressive reconnection that continuously forms flare loops.
From the flare ribbon evolution, we can measure the reconnection rate
by summing up magnetic flux in newly brightened UV pixels \citep{fletcher01,qiu2002,qiu2004,saba06,qiu10}.
An automated procedure has been developed to identify flaring pixels and minimize measurement uncertainties
caused by fluctuations in photometry calibration and by non-flaring signatures \citep{qiu10}.
We use the SDO/HMI LOS magnetogram combined with the SDO/AIA 1600 \AA \ \ images of flare ribbon evolution
(see Figure~\ref{fig3}) to derive the magnetic reconnection flux and magnetic reconnection rate,
which are displayed in the middle and bottom panels of Figure~\ref{fig4}, respectively.
The calculation is made between 11:00:00 UT and 13:20:00 UT.
In this flare, the uncertainty in reconnection flux measurement is about 40\%, mostly
caused by the imbalance between the measured positive and negative fluxes, although
the two fluxes well track each other during the flare evolution.
At 11:10~UT when ribbon brightening starts, the reconnection flux starts to rise, and the
total reconnection flux is measured to be 1.8 $\times$ 10$^{21}$ Mx. The reconnection rate is
the time derivative of the reconnection flux, which amounts to 1.5$\times$ 10$^{18}$ Mx s$^{-1}$ at around
11:30~UT.

The associated CME is a partial halo CME observed by LASCO with an average speed
of 736 km s$^{-1}$ in the LASCO C2 and C3 field of view, according to the LASCO CME catalog
by the Coordinated Data Analysis Workshops (CDAW)\footnote{\url{http://cdaw.gsfc.nasa.gov/}} CME
catalog \citep{Gopalswamy2009}. The CME event, particularly in its early stage, is best observed
by the STEREO instruments from the limb. Figure~\ref{fig2} shows the CME evolution observed by EUVI, COR1
and COR2 onboard STEREO. The CME is first seen at 11:06:25~UT as a faint loop structure in the running
difference images at 195\AA\ by EUVI-B, and expands afterwards. The bright CME front
is clearly seen in COR1 at 11:30:55 UT. Five minutes later, a bright core beneath the
front also appears in the COR1 field of view. These structures can also be identified in
COR2 images from 12:24:55 UT.

To measure the height of the CME,  we track the position of
the leading edge of the CME along a guide line, indicated in the figure,
in the direction of its ejection. The CME leading edge in best visible in
the running difference images by EUVI, COR1, and COR2.
This position is tracked manually several times. The mean and standard deviation
of these measurements are presented in the figure as the CME height and its uncertainty,
respectively. The middle panel of Figure~\ref{fig4} shows the measured
height (starting from the solar center) of the CME front together with the reconnection flux measured from the flare ribbon
evolution. We derive the CME velocity using a linear fit
to every four consecutive points, in order to reduce large fluctuations in the original measurements.
The uncertainty indicated in the velocity plot is the standard deviation of this linear fit.
The plot shows that the CME starts moving slowly at a speed below 100 km s$^{-1}$ when flare UV
emission starts, and then speeds up in the next 10 minutes. It achieves the maximum velocity
of about 1000 km s$^{-1}$ around the time of the GOES soft X-ray emission peak, when the increase of the
reconnection flux starts to slow down substantially. The CME then moves at a relatively constant speed of about
700 km s$^{-1}$ for the next few hours. The CME velocity profile shows a temporal correlation with the
reconnection flux as well as the GOES soft X-ray light curve during the rise
phase, consistent with previous observations of the timing relationship between fast CMEs and
flares \citep{zhang01,qiu2004,patsourakos10,patsourakos13,cheng14}.
The CME acceleration is also estimated and compared with magnetic reconnection rate. The bottom
panel of Figure~\ref{fig4} shows that these two quantities exhibit similar evolution trends. The
CME acceleration peaks around 11:30 UT with a peak value of 0.4 km s$^{-2}$, and
diminishes after 12:00~UT, when flare reconnection has also stopped. At these times when acceleration/reconnection
starts (11:10~UT), peaks (11:30~UT), and stops (12:00~UT), the front of the CME is at 0.2, 1, and 3 solar radii
above the surface, respectively.

In summary, these observations show the following stages of the CME and flare evolution:\\
 11:10 start of flare ribbon emission, CME motion, and reconnection;\\
 11:20 start of SXR rise, significant CME acceleration, and fast reconnection; \\
 11:30 peak of CME acceleration and reconnection rate;\\
12:00 peak of SXR emission and CME velocity, and reconnection diminished. \\

In the standard flare-CME model, energy stored in the magnetic
field is released by magnetic reconnection to heat flare plasmas,
as well as to accelerate the CME.  The observed coincidence between the CME evolution
and magnetic reconnection, within the uncertainty of a few minutes, suggests that
acceleration of the CME is related to the reconnection process.

\subsection{EUV Dimming}
\subsubsection{Temporal and Spatial Evolution}
Coronal dimming is often observed to accompany CMEs. In this event, dimming signatures are observed
in AIA images of several EUV bands, most prominently 171, 193, 211, and 335
bands, suggesting that the characteristic temperature of the dimming areas ranges between 1 to 3 MK.
By checking the full disk images, the most significant dimming occurs near the flare region.
No other obvious dimming is detected during that time.

In the middle panel of Figure~\ref{fig4}, we plot 171 and 193 light curves produced from
pixels exhibiting dimming. The pre-dimming counts level in these pixels is
within $I_q \pm \sigma$, $I_q$ and $\sigma$ being the median counts and standard
deviation, respectively, of a control region free of coronal loops, and EUV counts
in these pixels keep decreasing continuously for a
few minutes. In this way, we exclude apparent dimming signatures due to removal or motion
of overlying coronal loops during the eruption \citep{Qiu2007, hock2010, harra11, Hock2012}.
The plots show that coronal dimming evolves along with the flare reconnection and CME eruption.
EUV emission in the 193 passband starts to decrease quickly at 11:15 UT, 5 minutes after
the start of the flare reconnection and CME takeoff, whereas dimming in the 171 band starts about 5
minutes later. The maximum dimming occurs around the peak time of the GOES SXR emission as well as the
CME velocity, and then EUV emission in these dimming regions starts to rise as post-flare loops form in
these regions.

\begin{figure}
\epsscale{0.8} \plotone{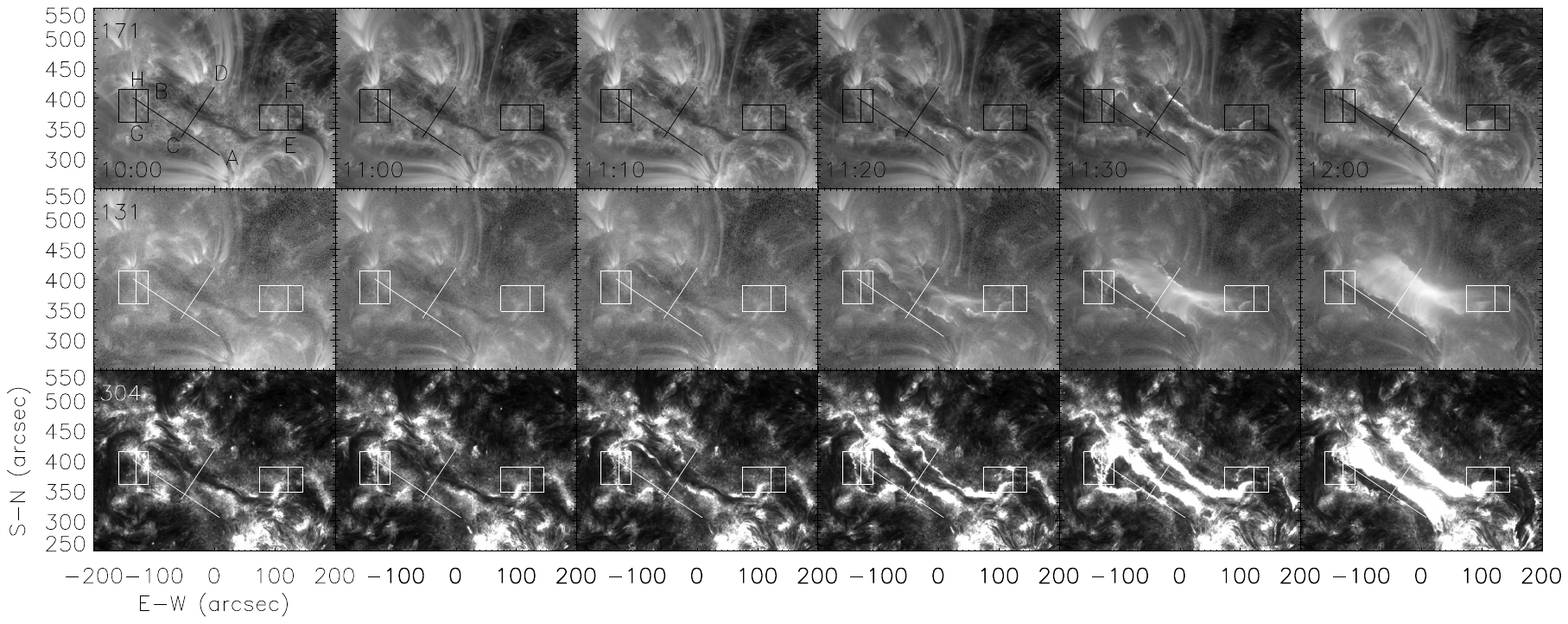}
\caption{Snapshots of flare and dimming evolution in three different passbands (171, 131 and 304 \AA\ \ ) at six times.
The boxes indicate regions of interest where we study dimming signature in details. The four slices sample
different dimming regions, and stack plots in the next figure are constructed along these slices to study tempo-spatial
variation of dimming. \label{fig5}}
\end{figure}

To study the evolution of dimming and its relation to the eruptive flare, Figure~\ref{fig5}
gives an overview of the flare region in three different passbands by AIA in 171, 131,
and 304 \AA\ at several critical times during the flare/CME evolution.
The dimming pixels selected above mostly locate in EUV moss areas adjacent to the flare ribbons.
One dimming region is found at the left end of the flare ribbon in positive magnetic fields. Dimming also
occurs at the right end of the conjugate flare ribbon in negative magnetic fields.
These two regions are indicated by the rectangular boxes in Figure~\ref{fig5}, as well as in
the magnetogram in Figure~\ref{fig3}. We also find a band of dimming region
along the flare ribbon in positive magnetic fields, indicated by the long parallelogram in Figure~\ref{fig3}.
These regions are denoted as ``LF" (left foot), ``RF" (right foot), and ``RB" (ribbon) in the figures as
well as the following text. Dimming is also visible along the other ribbon in negative
magnetic fields, but not as outstanding and widespread as in the RB region, perhaps because
of projection effects given the disk position of the active region. Seen in the leftmost panels of
Figure ~\ref{fig5}, before the flare, these regions are characterized by EUV moss structures
at the base of the corona, and are mostly free from overlying coronal loops. After the eruption,
the moss structure in some areas is removed, causing coronal dimming seen in a few EUV bands including the 304\ \AA\ band
observing the transition region, as shown in the rightmost panels of the figure.

\begin{figure}
\epsscale{0.8} \plotone{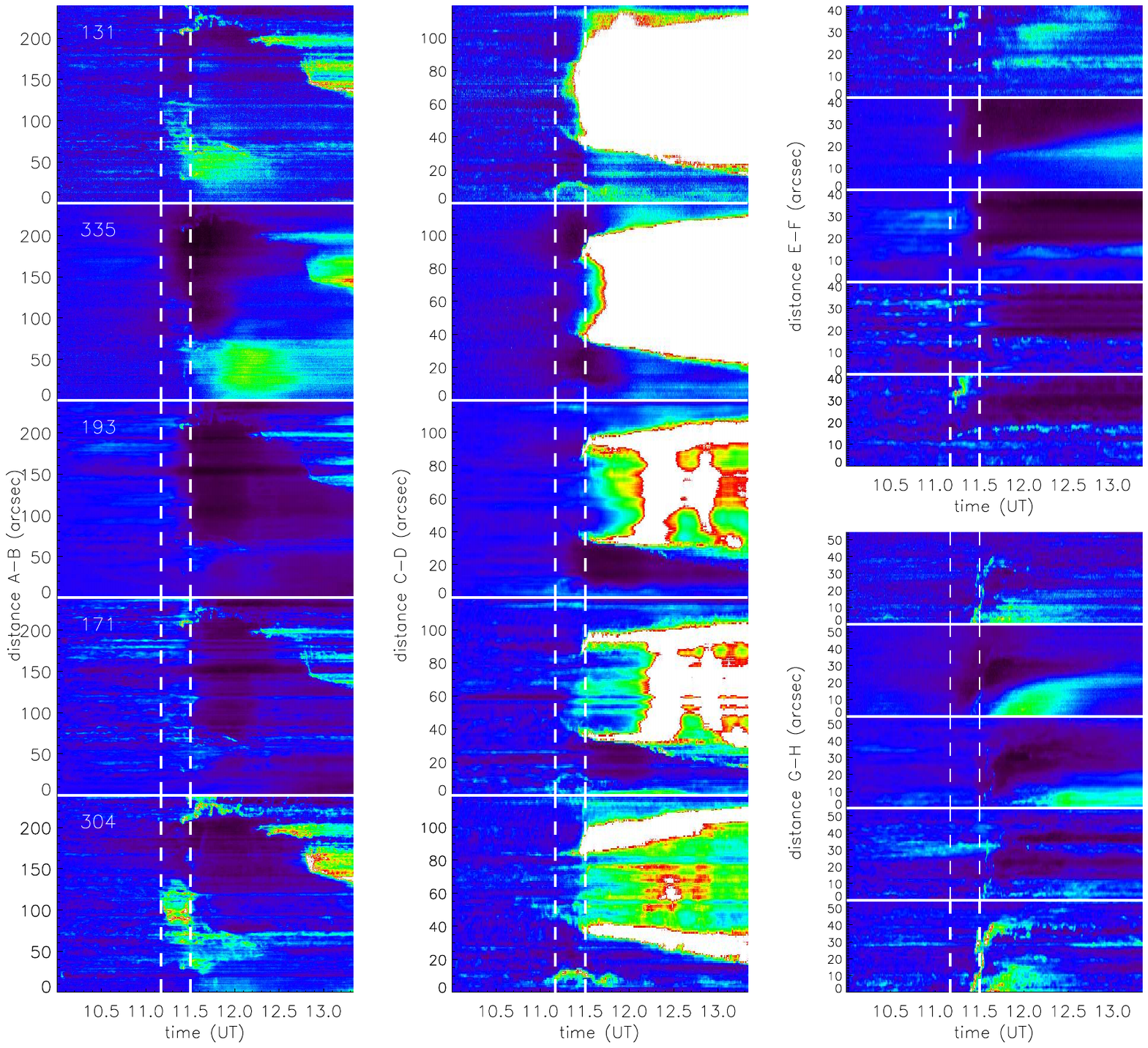}
\caption{Dimming evolution along the marked slices in Figure 5. Two vertical lines indicate times at 11:10 and 11:30~UT, respectively.
Dimming occurs in multiple EUV bands, and major dimming starts after 11:10 UT.  \label{fig6}}
\end{figure}

To analyze the temporal and spatial pattern of dimming evolution, we sample pixels in these regions by generating
stack plots of EUV emissions in a few passbands along a few slices across
these regions. These slices are marked in Figure~\ref{fig5}. The left and middle panels in Figure~\ref{fig6} show
the stack plots of EUV counts along slices AB and CD, which are placed along and across the dimming band RB next
to the flare ribbon, respectively. The right panel shows the stack plots along slices EF and GH, which
sample the two dimming regions at the ends of the conjugate ribbons. From top to bottom, the plots
are generated for the AIA 131, 335, 193, 171, and 304 \AA\ passbands, sampling plasma emission at different
temperatures. Each stack plot is normalized to the EUV counts at 10:00:12~UT. The two vertical dashed lines
indicate the times at 11:10~UT and 11:30~UT, respectively, when CME acceleration and flare reconnection start and peak.

Seen in these plots, dimming is observed in multiple EUV bands including the EUV 304\ \AA\ band. In most dimming
pixels, significant dimming starts after the flare/CME onset (11:10 UT) in the 335 and 304 passbands, and even
later in other wavelengths. In a few locations, gradual dimming appears to occur before 10:30 UT.
The middle panel of Figure~\ref{fig6} sampling the evolution across
the flare ribbon along slice CD also shows apparent spreading of dimming away from the PIL, followed by spreading of ribbon
brightening. For example, seen in the 193\ \AA\ passband, between 11:25 and 11:35 UT, the
dimming front along the ribbon spreads away from the PIL at the speed of about 10 km s$^{-1}$.
On the other hand, the right panel shows that dimming along the EF and GH slices across the
LF and RF regions also spreads out, but these pixels are not brightened later on.

The temporal and spatial evolution of dimming in multiple wavelengths suggests that dimming in these regions
is not caused by thermal dimming or disruption of pre-flare active region loops \citep{harra11,hock2010}, neither
do they appear to be obscured by absorptive materials. Most likely, dimming is caused by density
decrease due to expansion of corona structures anchored at these locations. Specifically,
the LF and RF regions at the two ends of the flare ribbons locate at oppositely curved magnetic
elbows that loop out on opposite ends of the neutral line \citep{sterling97,moore01}, and likely
map the footpoints of the erupting CME, similar to the event reported by \citet{webb00}.
On the other hand, the narrow band of dimming along the outer edges
of flare ribbons is brightened afterwards.  The morphology evolution at these locations
much resembles the scenario depicted by \citet{forbes00,svestka96} and \citet{moore01}, where
the erupting flux rope stretches the overlying coronal arcade causing dimming at the feet of the
arcade; the stretched field lines then reconnect and form closed loops, and therefore these
locations are brightened due to heating along closed loops. Note that at these locations,
reconnection as inferred from the ribbon brightening starts after the dimming; yet in other
regions, such as in the inner part of the flare ribbons, reconnection takes place much
earlier, and dimming is not observed in some of these regions. As a result, the global
signature of significant dimming appears to occur after the onset of reconnection.

\subsubsection{Coronal Dimming versus CME Evolution}
To further examine the timing of coronal dimming with respect to the CME eruption,
we analyze EUV light curves of the three regions, with a focus on the 193, 171, and 304 passbands
that observe the base of the corona. The AIA response functions for the 193 and 171 bandpasses each
have a dominant single peak at temperatures below 10~MK \citep{Boerner2012}, so observations in these passbands
provide a more reliable estimate of the characteristic temperature of the dimming plasma. Light curves
of the 211 passband are very similar to those at the 193 passband and are therefore not presented.
Emission in the 304~\AA\ passband includes a large contribution of the He {\sc ii} 304~\AA\ line that
forms in the transition region or upper chromosphere at a characteristic temperature of about 0.1~MK.
Traditionally, dimming is not often studied in this wavelength, but we present observations in this wavelength
as a reference. Observations show that dimming is also visible in the 304 band, likely caused by the
diminished transition-region emission due to decreased coronal pressure along expanding field lines anchored at these locations.

In the top row of Figure 7, we plot light curves of the total EUV counts from all pixels in LF (left),
RF (middle), and RB (right) regions, respectively. Each light curve is normalized to the
pre-eruption counts at 10:00~UT. In the second row, the inverse of the same light curves
are presented, in order to compare with the CME height profile overplotted in symbols.
In these light curves, the 171 total counts appear to be rising (brightening)
rather than decreasing (dimming), partly because not all pixels in the regions are dimming.
We select pixels that show evidently and continuously decreased counts, or dimming, in
each bandpass during the eruption, and compare their light curves with the CME height profile.
We also find that, in the 171 passband, a significant number of selected
dimming pixels in all three regions exhibit a brief brightening before dimming.
This will be discussed later. Light curves from two groups of dimming pixels
are presented in the following four rows of the figure.

In the third and fourth rows, we show light curves of the total counts of the first group of pixels that exhibit
dimming in both the 193 and 171 bandpasses without brightening before dimming. In the bottom two rows,
we plot light curves of the second group of pixels with dimming in both bandpasses but also significant
brightening in the 171 passband before dimming. All these light curves show a rapid decrease of EUV counts
in the 193 and 171 bands at or after the onset of fast CME acceleration at around 11:10~UT. Dimming well
tracks the CME height for about 20 minutes, and the slope of the dimming growth, namely the rate of the counts
decrease, particularly in the 193 bandpass, is comparable with the growth rate of the CME height. When the
two groups of the dimming light curves are compared, it appears that the dimming slope of the 193 light curve
in the second group, which show significant brightening in the 171 band before dimming, is slightly larger
than in the first group. Finally, in some regions and at some wavelengths, gradual dimming starts earlier, well
before 11:00~UT, consistent with the stack plots in Figure~\ref{fig6}.

If coronal dimming results from the decreased density in expanding coronal structures, the dimming light curve at a given
location is related to the height of the structure anchored at that location. It is not possible to identify
expanding coronal structures anchored at different locations or pixels of dimming; however, we may use
the measured CME height as a common reference to examine the growth of dimming in various places.
For this purpose, we make a routine to select and fit the dimming light curve in each of the three bandpasses (193, 171, and 304)
in each dimming pixel as related to the CME height by the following function
\begin{equation}
{\rm ln} \left(\frac{I_0}{I}\right) \sim \alpha {\rm ln} \left(\frac{H}{H_0}\right)
\end{equation}
where $I$ refers to the EUV counts in a given pixel observed by AIA, and $H$ is the CME height above
the solar surface measured in STEREO observations. $I_0$ and $H_0$ are values at 11:06~UT, when the
CME is first visible in STEREO observations. The fitting parameter $\alpha$ gives the slope, or the growth
rate, of dimming relative to the growth of the CME height, both of which are interpolated to a grid of
1-min time cadence. For each pixel, fitting is performed in the time range when
EUV counts monotonically decreases (i.e., dimming). The period of pre-dimming brightening is excluded
from fitting, and obviously the gradual dimming period before CME is visible is also excluded. In some pixels,
post-eruption dimming apparently evolves in two stages, with a rapid first-stage dimming followed
by a gradual dimming. In this case, fitting is performed in two stages, and we use $\alpha$ from the first stage.

Figure~\ref{fig8} shows the statistics of the fitting parameter $\alpha$ for a few thousand
dimming pixels in the three regions. The fitted slope mostly ranges between 0.4 - 1.2, although
different regions show small variations. Goodness of fitting is measured by $\chi ^2$, and we
accept fitting results with $\chi ^2 \le 2$. The uncertainty in the fitting parameter $\alpha$
is about 0.2. For the LF region (left panel), about 75\% pixels show dimming in 193 \AA\ ,
25\% pixels show dimming in 171 \AA\ , and 20\% pixels show dimming in 304 \AA\ , the median
of the dimming slope ($\alpha$) being 0.45, 0.36, and 0.39, respectively. For the RF region (middle panel),
about 85\%, 36\%, and 39\%  pixels show dimming in 193 \AA, 171 \AA. and 304 \AA\ ,
the median of the slope being 0.5, 0.4, and 0.3 for the three passbands. For the RB region along
the flare ribbon (right panel), 73\%, 39\%, 36\% pixels show dimming in 193, 171, and 304 \AA\ , and
the median slope is 0.7, 0.5, and 0.3 for the three bands.
Pixels with a larger slope usually exhibit stronger maximum dimming (i.e., lower counts). We also find that
the dimming slope in the second group of pixels, which exhibit pre-dimming brightening, is statistically
larger than in the first group.

We further study the spatial distribution of the fitted dimming slope by mapping $\alpha$ in Figure~\ref{fig9}.
Different colors indicate different $\alpha$ values, which are indicated in the figure. The fitted results for the three
wavelengths (193, 171, and 304 \AA\ ) are displayed in the figure. We find more dimming pixels in 193 \AA\ than in
the other two wavelengths. A large percentage of pixels have a dimming slope between 0.4-1.2. A very small
number of pixels have $\alpha \ge 1.6$. Generally, the slope in most pixels is less than 1.2, both in
foot point regions and along the flare ribbon. Interestingly, the map shows a certain pattern of the
slope distribution, which is most evident in the 193 band. The right and left footpoint regions each
exhibit a core of a large slope (i.e., deep dimming) shown in cyan and green colors, surrounded by outer
layers in yellow and then red indicating dimming slope successively decreasing outward. The RB area also
shows a regular pattern: from right to left as approaching the left foot along the
ribbon, the slope becomes larger, from red ($\alpha$ = 0-0.4) to green ($\alpha$ = 0.8-1.2). Such a spatial pattern is
also seen in the other two bandpasses, but it is not as prominent as in the 193 band.

In summary, we find that the dimming slope mostly ranges between 0.4-1.2, and its spatial
distribution exhibits deep dimming cores, most evident in the 193 band, around the regions which probably map
the CME feet. These results will be discussed in the following section in the context of the CME expansion process.

\subsection{Coronal Dimming and CME Expansion}
The fact that dimming curves well track the CME time height profile may indicate that dimming
is related to the expansion of the CME. As the coronal structure expands, density decreases, so
the emission measure along the line of sight decreases to produce diminished EUV emission.
In the following, we test a few CME expansion models to investigate whether these models
can explain the observed dimming signatures.

The AIA observed EUV emission in terms of data counts in a given pixel can be approximated by
\begin{equation}\label{1}
    I \approx n^{2} R(T) L,
\end{equation}
where $L$ is the length along the line of sight, $n$ and $T$ are the mean electron density and
temperature along $L$, and $R(T)$ is the temperature response function of the AIA filter.

If the CME expansion is one-dimensional, for mass conservation, the plasma density inside the
expanding coronal structure decreases linearly with the increasing $L$, namely $n \propto L^{-1}$.
If we assume $L \sim H$, $H$ being the CME height, for an isothermal model where plasma temperature $T$ is constant,
the observed EUV counts variation is related to the CME height by
\begin{equation}\label{1}
    \frac{I}{I_0} = \frac{n^2 R(T)H}{n_0^2 R(T)H_0} = \frac{H_0}{H},
\end{equation}
where $n_0$, $H_0$, and $I_0$ are initial values at 11:06 UT, when CME is first visible in the STEREO
field of view. The CME height $H$ is measured above the solar surface.

If the expansion is self-similar, the plasma density decreases with the increasing
$L$ by the relation $n \propto L^{-3}$. Again, assuming that $L \sim H$, in
an isothermal model, we arrive at
\begin{equation}
\frac{I}{I_0} = \left(\frac{H_0}{H}\right)^5.
\end{equation}

If on the other hand, the CME structure undergoes adiabatic expansion, then $TV^{\gamma-1} = const.$, $V$ being the plasma volume,
with $\gamma = \frac{5}{3}$. For the one-dimensional expansion $V \propto L \propto n^{-1}$. Therefore,
\begin{equation}
T = T_0\left(\frac{L_0}{L}\right)^{2/3},
\end{equation}
where $L_0$ and $T_0$ are initial values at 11:06~UT. With $L \sim H$, the observed EUV counts change as
\begin{equation}
\frac{I}{I_0} = \frac{R(T)}{R(T_0)} \frac{H_0}{H}.
\end{equation}

For an adiabatic self-similar expansion, we have
\begin{equation}
T = T_0 \left(\frac{L_0}{L}\right)^{2},
\end{equation}
and
\begin{equation}
\frac{I}{I_0} = \frac{R(T)}{R(T_0)} \left(\frac{H_0}{H}\right)^5.
\end{equation}

With these models, we estimate the observed dimming evolution ($I/I_0$)
as related to the measured height $H/H_0$ of the CME. First, applying the filter ratio method
to AIA data at 171 \AA, 193 \AA, and 211 \AA, we find the plasma
temperature $T_0 = 1.4$~MK at 11:06~UT, which is a reasonable active region
coronal temperature. In the isothermal model, $T = T_0 = 1.4$~MK.
In the adiabatic model, $T$ varies as described by Eq. (5) and (7) for the
one-dimensional and self-similar expansion, respectively. As the CME propagates, the derived
temperature decreases to 10$^{5.5}$ K or even lower. Then from Eqs. (3), (4), (6) and (8),
we calculate the EUV light curves $I/I_0$ as would be observed in the AIA 171 and 193 passbands for the
four different CME expansion models. From the analysis above, if the CME expansion is isothermal,
the EUV counts monotonically decrease as the CME expands by $I_0/I \sim (H/H_0)^{\alpha}$, and
the power index $\alpha$ should be 1 for the one-dimensional expansion or 5 for the self-similar expansion
in both the 193 and 171 bands. On the other hand, if the CME undergoes an adiabatic expansion, as density decreases,
the plasma temperature also decreases, so the EUV counts variation becomes a bit complicated. The
response function for the 193 band peaks at around 1.5~MK, comparable with $T_0$, so adiabatic expansion
will reduce the observed 193 counts further more than in the case of isothermal expansion. Therefore,
the predicted slope of the 193 dimming is greater than 1 in the one-dimensional expansion model and greater than 5
in the self-similar expansion model. In the 171 passband, the filter response function peaks
at 0.8~MK, lower than $T_0 = 1.4$~MK, hence the predicted light curve for adiabatic expansion models
is more complicated.

Figure~\ref{fig10} shows these modeled light curves in the 193 (left) and 171 (right) bands
using Eqs 3, 4, 6, 8, in comparison with some observed light curves in the left footpoint
region (top), right footpoint region (middle), and ribbon region (bottom).
As discussed above, in the 193 band, the predicted light curves with adiabatic expansion models (blue and red)
have a larger slope than isothermal expansion models (cyan and green), and in the 173 band, adiabatic
expansion produces apparent brightening in the observed light curves, followed by a sharp dimming with a large
dimming slope. As shown in the last section, the observed EUV light curves have the dimming slope $\alpha$
mostly ranging between 0.4 and 1.2, whereas the predicted EUV light curves should have slopes greater than 1.
Therefore, dimming in many pixels do not agree with any of the expansion models. In the figure, we
choose to plot light curves of the pixels in dimming cores, which have relatively large slopes.
For the three selected regions (LF, RF, RB), the 193 counts variation of these pixels most closely follows
the 1d isothermal expansion model. In the 171 band, most of these same pixels appear to be brightened first,
similar to the 1d adiabatic curve, and then the subsequent dimming curve again mostly resembles the 1d
isothermal expansion model.

From the fitting result, we can exclude the self-similar expansion process for this event, if expansion of
the coronal structures anchored at the dimming locations can be approximated by, or linearly scaled with,
the measured CME height. The core regions of deep dimming exhibit EUV light curves mostly tracking the
1d isothermal expansion model based on the measured CME height. It is not clear whether the
observed initial brightening in the 171 band is caused by adiabatic expansion for the initial period,
since light curves in the 193 band of the same pixels during that period do not follow the prediction
by the adiabatic expansion model.

We can also estimate the mass and kinetic energy of the erupting CME,
assuming that dimming is primarily caused by expansion of CME field lines anchored at these
dimming locations. From the filter-ratio
analysis mentioned above, we get the pre-eruption mean plasma temperature
around 1.4 MK. The pre-eruption mean plasma density is estimated by
\begin{equation}
n_0 = \left[\frac{I_0}{R(T_0)L_0 }\right]^{\frac{1}{2}},
\end{equation}
and the total mass of the CME is
\begin{equation}
m = \sum (n_0 {\Delta} S f_1 L_0),
\end{equation}
where $\Delta S$ is the area of a dimming pixel, and $f_1$ is a scaling
factor larger than 1, since
the half length of the structure is usually a few times the length scale along the
line of sight $L_0$.  We may further write $L_0 \approx f_2 H_0$, with $f_2 < 1$, indicating
that $L_0$ is a fraction of the CME height $H_0$. Combining the above equations, the CME mass is given as
\begin{equation}
m \approx \left[\frac{I_0}{R(T_0)}\right]^{\frac{1}{2}} f_1 \sqrt{f_2 H_0}\sum  {\Delta} S
\end{equation}
The values of $f_1$ and $f_2$ much depend on the geometry, including the aspect ratio
of the CME structure. Since the CME structure before its eruption is hardly visible in the corona,
its unknown geometry presents the largest uncertainty in the mass estimate.
If the pre-eruption CME structure is approximated by a semi-circular loop with the axis
of the loop in a plane perpendicular to the solar surface, and the major radius of
the loop, which is the same as $H_0$ in this geometry, is 4 times the minor radius of the
loop, then $f_2 \approx 0.5$, and $f_1 \approx \pi$. With this geometry, the mean density is estimated to
be $\sim$ 1.0 $\times$ $10^9$ cm$^{-3}$. By summing up all the dimming pixels, the total mass loss $m \approx 1.1
\times$ 10$^{15}$ g. The derived maximum CME velocity is about 1100 km s$^{-1}$, yielding the
maximum kinetic energy 6.9$\times$ 10$^{30}$ ergs. If we consider that dimming along the other ribbon in the
negative field is obscured by projection effects, the mass and kinetic energy could be
increased, but by no more than a factor of two. In comparison, the CME mass and kinetic
energy are estimated to be 4.3$\times 10^{15}$ g and 1.5$\times 10^{31}$ ergs (using the mean
speed of 736 km s$^{-1}$), respectively, with the LASCO white-light coronagraph observations (CDAW). Estimate
from the dimming analysis appears to set a lower-limit of the CME mass,  which is about one-quarter of
the estimate from coronagraph observations. Such an estimate is very rough, subject to the
uncertainties in $f_1$ and $f_2$ because of the unknown CME geometry; it is therefore
only appropriate as an order-of-magnitude estimate.

Finally, we measure the magnetic flux encompassed by the dimming regions. In the entire active region,
about 4\% pixels exhibit dimming in both the 193 and 171 bands with the longitudinal
magnetic flux density greater than $\pm$ 10~G. The sum of the magnetic flux in these pixels
amounts to 3.4$\times 10^{20}$ Mx in positive fields and 1.4$\times 10^{20}$ Mx in negative fields.
Magnetic field in dimming pixels in the LF and RB regions is predominantly positive, and field in the RF
region is negative. That the negative flux is smaller than the positive flux is due to lacking dimming
signatures along the flare ribbon in negative fields most likely because of projection effect.
The total flux at the feet of the expanding field lines should be closer to the measured positive flux,
which is about 20\% of the total reconnection flux measured from flare ribbons.

\section{DISCUSSIONS AND CONCLUSIONS}
The analysis presented in this paper is focused on EUV
dimming associated with an eruptive flare. Coronal dimming is
observed near flare ribbons in multiple EUV passbands of AIA,
together with a fast CME observed by STEREO. Significant dimming
starts at or after the onset of CME eruption or flare reconnection, and the dimming depth
(decrease of EUV counts) grows as the CME propagates
outward. The time sequence of the flare/CME/dimming event in the
early phase of eruption is as follows:
(1) 11:06 CME first seen by EUVI/STEREO; (2) 11:10 start of CME motion and reconnection;
(3) 11:20 start of significant CME acceleration, fast reconnection, and rapid dimming; (4)
11:30 peak of CME acceleration, reconnection rate, and strong EUV dimming (5)
12:00 peak of CME velocity, GOES emission, and EUV dimming, whereas reconnection diminished.
Such timing among these events indicates that fast reconnection, CME expansion, and coronal dimming
are intimately related.

The spatial distributions of the dimming signatures are also studied.
Dimming is primarily located near flare ribbons. Two dimming regions
are found at the opposite ends of the two ribbons, which likely map
the conjugate feet of the CME. During the flare eruption, dimming
along the outmost edges of one flare ribbon is also observed, which spreads outward
at an apparent speed of $\sim$ 10 km s$^{-1}$. It is followed
by brightenings of flare ribbons. This signature agrees with the standard
model, that overlying coronal arcade is stretched by the erupting CME causing
coronal dimming at the feet; the arcade then reconnect and produce brightening
ribbons. Furthermore, we find that dimming regions exhibit deep dimming cores, where
dimming grows more rapidly, and the dimming growth becomes smaller outside the cores.
These signatures are similar to the MHD simulation results by \citet{Downs2015}.

Our analysis of the EUV dimming in multiple wavelengths suggests that
dimming in this event is most likely caused by plasma density depletion when
overlying coronal structures expand. Assuming that the expansion of these structures anchored
at the dimming locations can be approximated by the CME height measured from STEREO observations,
we predict dimming light curves using four simple models, i.e., one-dimensional adiabatic,
self-similar adiabatic, one-dimensional isothermal, and self-similar isothermal expansion models.
Observed light curves in thousands of dimming pixels are compared with the model predicted light curves
in the AIA 193 and 171 bands. We find that the dimming variation in the deep dimming cores
well follows the predicted light curve by the 1d isothermal expansion model, in which case,
the dimming variation is linearly scaled with the CME height. Dimming light curves outside
the cores do not agree with any expansion model.

From the dimming analysis, we have estimated the mass and kinetic energy of the erupting CME, as well
as the magnetic flux at the feet of expanding field lines. These measurements
may set the lower-limits of relevant quantities of the CME.
It is possible that a bulk of the CME is formed by reconnection within a pre-flare
sheared arcade, whose flux and mass are entrained into the CME. If a major part of this
arcade does not rise or expand rapidly before reconnection, dimming signatures
would not be produced at the feet of these arcade field lines, and therefore the
mass and flux involved in these structures would not be counted in the dimming analysis.

Admittedly, our analysis is subject to over-simplified assumptions. We have assumed
that the expansion of coronal structures anchored at various dimming locations is
approximated by or linearly scaled with a single CME height measured in the STEREO FOV,
and a mean pre-eruption temperature is used for all dimming pixels. In reality, different
regions may be subject to different physical conditions. Nevertheless, the observed timing
relation among the flare, CME, and dimming, and the spatial pattern of dimming signatures can
provide useful constraints to advanced models studying
reconnection and eruption in solar corona. We also remark that gradual dimming is observed
at some locations, starting about 30 minutes before the onset of CME eruption and flare
reconnection. It is plausible that this gradual and weak pre-eruption dimming may reflect
slow expansion of coronal structures, which can be further studied in search for early signatures
of CMEs.

\acknowledgments {We thank the referee for constructive comments that help improve the clarity
of the paper. This work is supported by NSFC under grants 11303073,11373023, 11133004 and
11103008. This research is supported by the strategic priority research program
of the Chinese Academy of Sciences (Grant No. XDB09000000). This research is also supported by the open project of Key Laboratory
of Solar Activity, National Astronomical Observatories, Chinese Academy of Sciences.
The AIA data used here are courtesy of SDO (NASA) and the AIA consortium.
STEREO is the third mission in
NASA's Solar Terrestrial Probes program. The SECCHI data are
produced by an international consortium of the NRL, LMSAL and NASA
GSFC (USA), RAL and University of Birmingham (UK), MPS(Germany),
CSL (Belgium), IOTA and IAS (France). }


\begin{figure}
\epsscale{0.8} \plotone{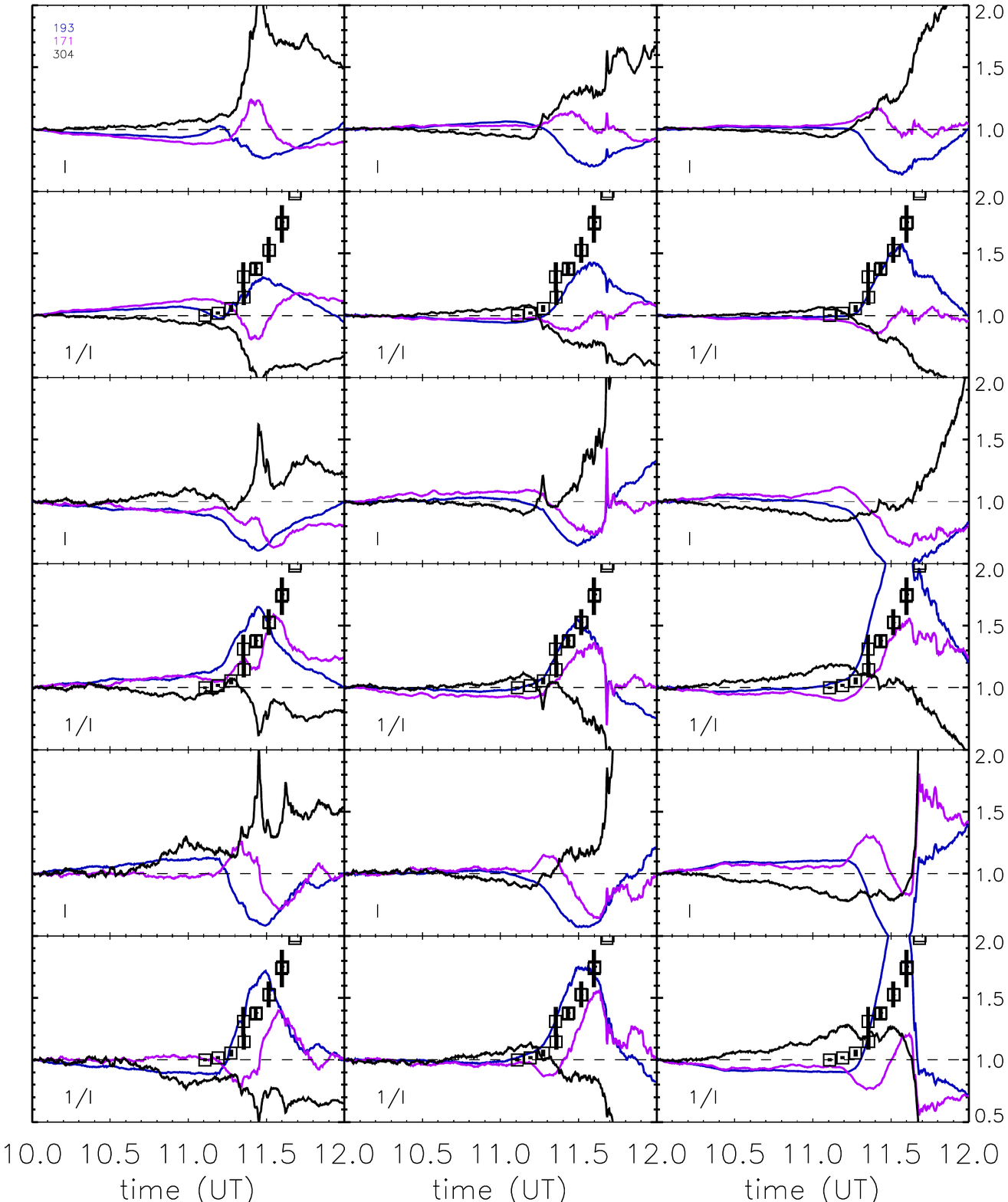} \caption{EUV light curves for the LF (left), RF (middle), and
RB (right) regions indicated in Figure 3. From top to bottom: the first two rows show light curves for all the pixels and its
inverse light curves, the middle two rows show light curves and their inverse for the pixels which show continuous dimming, and the
bottom two rows show light curves in dimming pixels which also exhibit brightening in 171 band before dimming. Symbols with error
bars show CME height measured from STEREO observations. \label{fig7}}
\end{figure}

\begin{figure}
\epsscale{0.8} \plotone{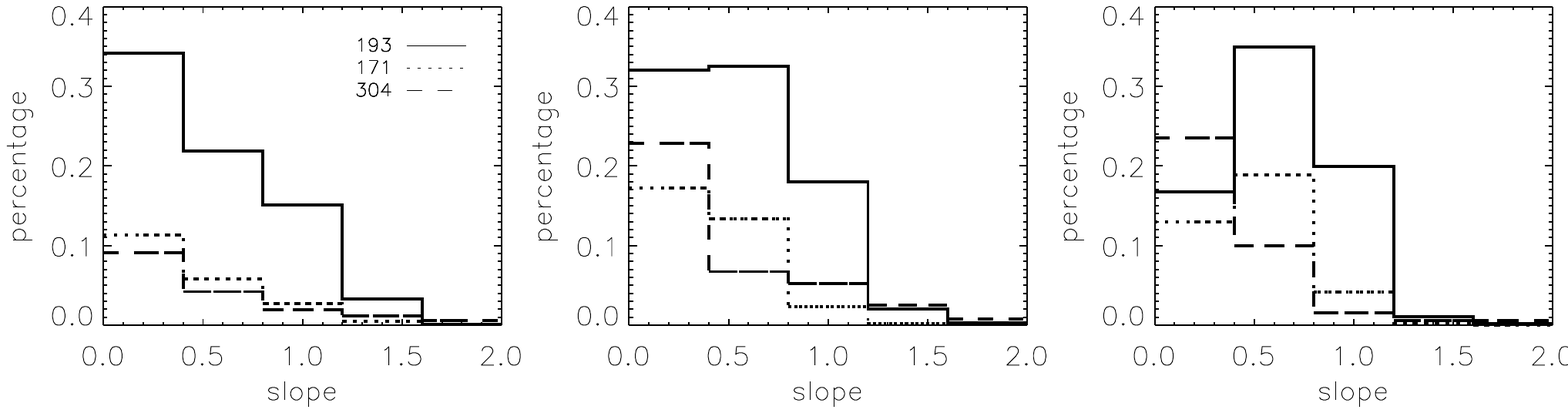} \caption{Histograms of the dimming slope $\alpha$ (see text) from fitting dimming light curves
in three regions, LF (left), RF (middle), and RB (right). The procedure of fitting is given in the text.
\label{fig8}}
\end{figure}

\begin{figure}
\epsscale{0.8} \plotone{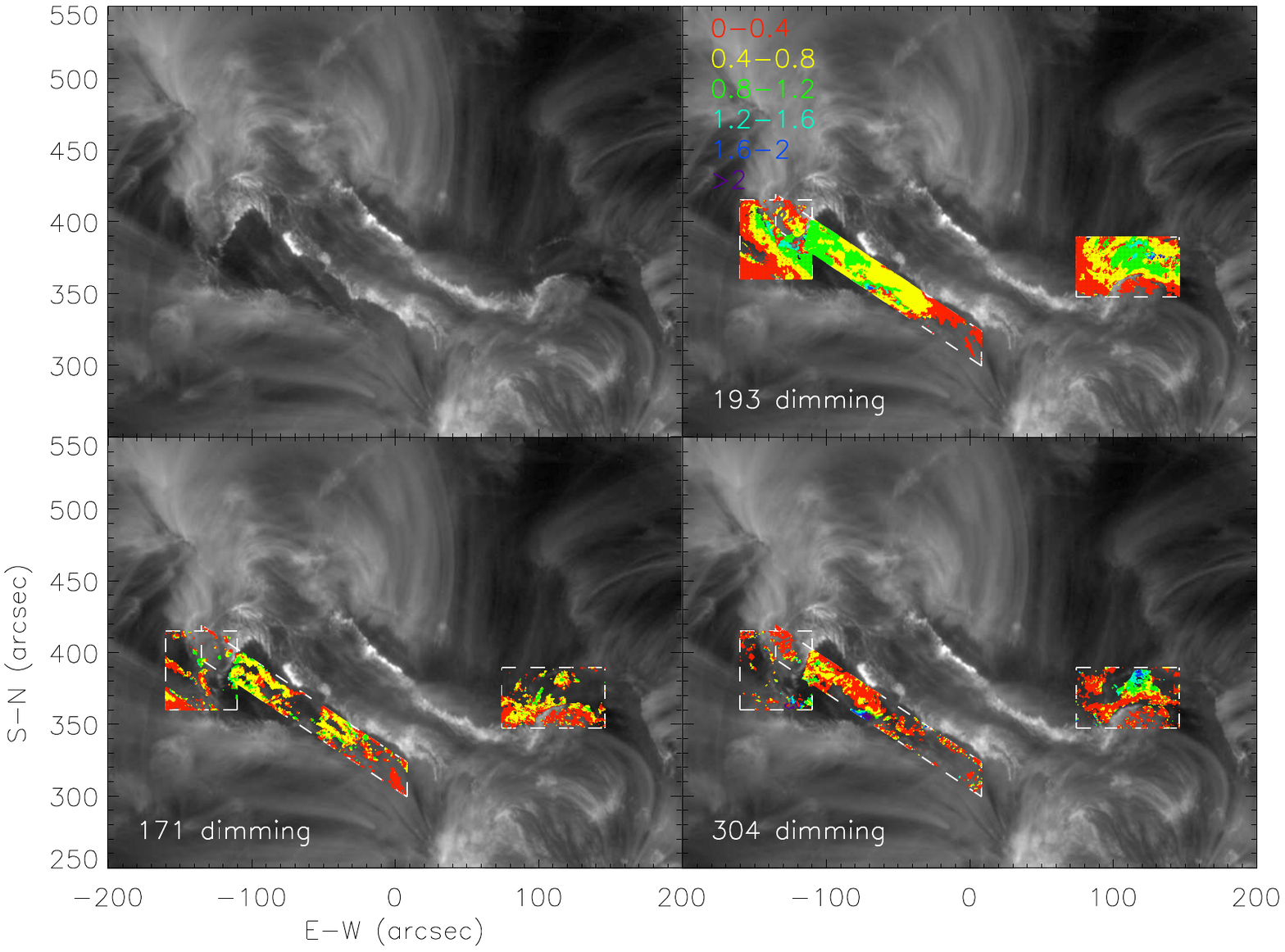} \caption{Spatial distribution of dimming slope $\alpha$ in three passbands.
It exhibits a regular pattern with deep dimming cores at the ends of the two ribbons. \label{fig9}}
\end{figure}

\begin{figure}
\epsscale{0.8} \plotone{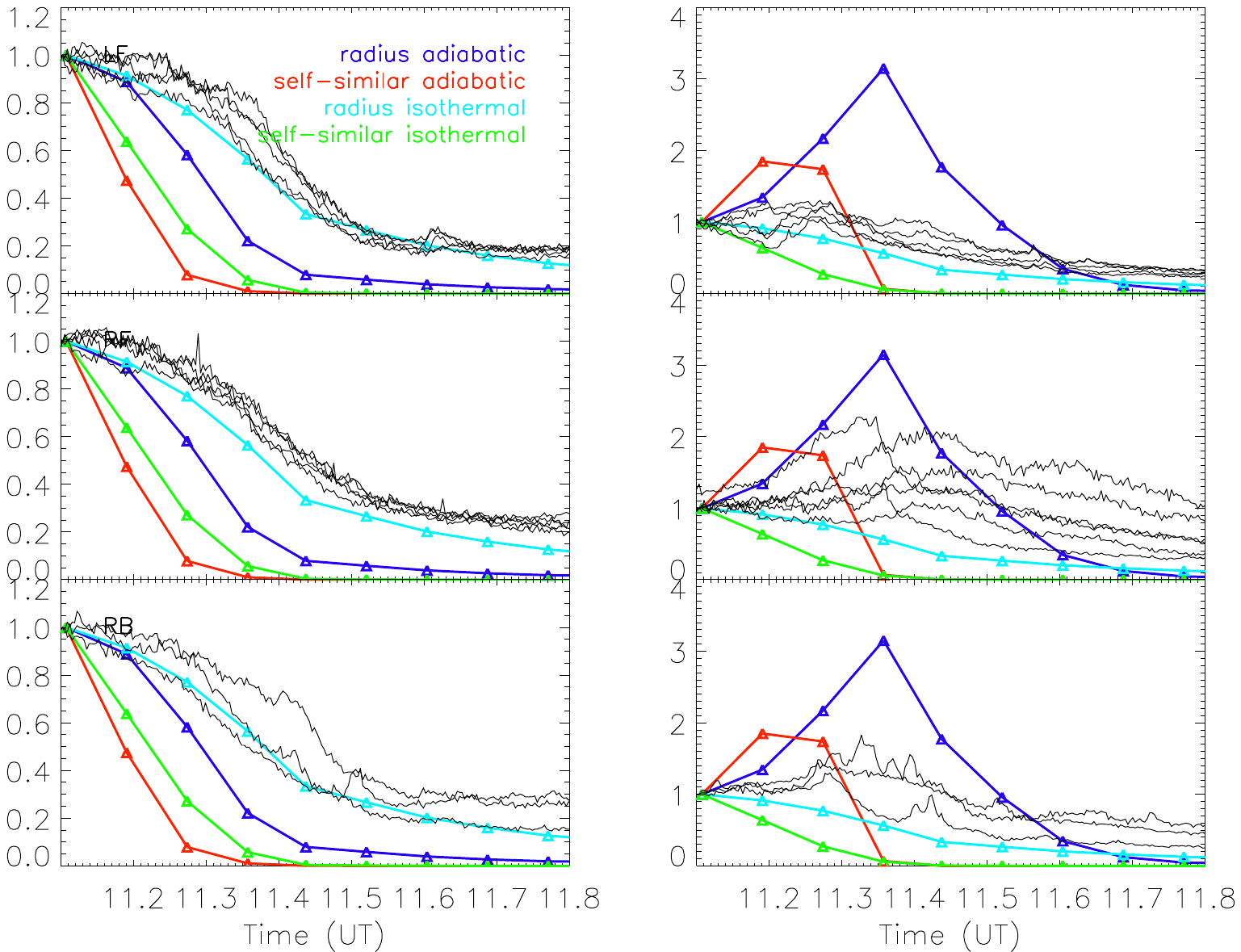} \caption{Left: predicted dimming light curves in 193 (left) and 171
(right) bands based on different CME expansion models (colorable lines), compared with observed dimming light curves from the dimming
cores in the LF (top), RF (middle), and RB (bottom) regions. \label{fig10}}
\end{figure}

\end{document}